\documentclass[aps,pra,twocolumn]{revtex4-1}

\usepackage{graphicx}
\usepackage{braket}
\usepackage{amsmath}
\usepackage{amssymb}
\usepackage{amsthm}
\usepackage{gensymb}

\DeclareMathOperator{\E}{\mathbb{E}}
\DeclareMathOperator{\G}{\mathrm{\Gamma}}

\begin{document}
\title{Turbulence-induced optical loss and cross-talk in spatial mode multiplexed or single-mode free-space communication channels}
\author{K. S. Kravtsov$^{1,2}$}\email{ks.kravtsov@gmail.com}\author{A. K. Zhutov$^3$} \author{I. V. Radchenko$^3$} \author{S. P. Kulik$^{3,4}$}
\affiliation{
$^1$ A.M.Prokhorov General Physics Institute RAS, Moscow, Russia \\
$^2$ National Research University Higher School of Economics, Moscow, Russia\\
$^3$ Quantum Technology Centre of Moscow State University, Moscow, Russia\\
$^4$ Faculty of Physics, M.V.Lomonosov Moscow State University, Moscow, Russia}

\date{\today}
\begin{abstract}
Single-mode or mode multiplexed free-space atmospheric optical channels draw increasingly more attention in the last decade. The scope of
their possible applications spans from the compatibility with the telecom WDM technology, fiber amplifiers, and modal
multiplexing for increasing the channel throughput to various quantum communication related primitives such as
entanglement distribution, high-dimensional spatially encoded quantum key distribution, and relativistic quantum
cryptography. Many research papers discuss application of specific mode sets, such as optical angular momentum
modes, for communication in the presence of atmospheric turbulence. At the same time some basic properties and
key relations for such channels exposed to the atmospheric turbulence have not been derived yet. In the current
paper we present simple analytic expressions and a general framework for assessing probability density functions of channel transmittance as well
as modal cross-talk coefficients. Under some basic assumptions the presented
results can be directly used for estimation of the Fried parameter of the turbulent channel based on the
measured statistics of the fundamental mode transmittance coefficient.
\end{abstract}
\maketitle

\section{Introduction}
Transmission of information along line-of-sight free-space optical channels is an important communication technique,
which was used by humanity for centuries. Automation of the originally manual information transfer, and introduction of
high-speed electronics and lasers led to many important practical applications in XX century. At the same time,
channel non-idealities, mainly, the atmospheric turbulence, became apparent. Turbulence effects were extensively studied in the
1970-s under the assumption of a broad, plane-wave-approximated light beam and a small point-like
photodetector~\cite{S1978}.

Later, the demand for {\em power efficient} free-space optical communication led to a more advanced model, where a single-mode Gaussian
beam from the source propagates toward a large aperture receiving telescope. In this model, turbulent effects shift and
distort the Gaussian beam, so it spreads in space and partially misses
the receiving aperture, thus producing the channel loss. This single-mode
transmitter and multimode receiver model is studied extensively in the series of more recent
papers~\cite{VSV12,VSV16,VSV17,VVS18}.

Following further technological advances, especially, breakthroughs in optical communications and quantum technologies,
the actual {\em single-mode} channel performance has to be studied. In this case the receiver collects only a particular
spatial mode, which is assumed to match the transmitted mode if there is no turbulence.
It is important, first, as a means of replacing
conventional single-mode fibers while keeping the infrastructure compatible with the WDM~\cite{YSJ07}, fiber amplifiers, coherent
modulation techniques~\cite{KRK18}, existing fiber-based quantum key distribution systems, etc.
Second, spatial mode-aware receivers allow for modal modulation, enabling data-efficient M-ary modulation formats,
not available in the fiber-based counterparts~\cite{KHF16}. Third, many independent data streams may be
spatially multiplexed into a single free-space channel, resulting in unprecedented data throughput at a single
wavelength~\cite{WYF12}. Finally, emerging quantum technologies reaching higher-and-higher quantum dimensionality, need corresponding
communication channels for exchange of such quantum states~\cite{NVG14,MMS15,SBF17}. The spatial degree of freedom may become the natural choice for
quantum computers talking to each other using high-dimensional spatial quantum states of photons~\cite{KHF15}.

The formulated problem of single-mode channel performance in the presence of atmospheric turbulence as well as related
questions of turbulence-induced modal cross-talk constitute the central question in this paper. By a single spatial mode we
understand an eigensolution of the propagation equation, so the mode remains itself while propagating through any distance.
Throughout the text we assume that if
the atmosphere was perfectly homogeneous and uniform, our optical system would be perfectly aligned without any
optical loss or cross-talk between the modes. The studied effects are solely due to the varying refraction conditions
in the turbulent atmosphere that distort the propagating modes, causing them to deviate significantly from the
unperturbed solution. We will introduce a framework that allows answering virtually any question about loss of power
in a particular mode or coupling of a particular mode into other modes, provided the turbulence parameters are known. We
show simple analytic solutions for the first-order approximation as well as numerically obtained results for higher order
approximations and compare them with experimental results. In particular, we show a very simple connection between the
transmittance statistics of a trivial single-mode fiber to single-mode fiber free-space link and the net turbulence strength
in this channel. This connection may be used for an easy parameter estimation either from the measured
channel statistics or from the known Fried parameter of turbulence.

\section{Turbulence model}

Turbulent phenomena in the atmosphere were first described by Kolmogorov~\cite{K41} in 1941, when he predicted
the scaling of a structure function proportional to $r^{2/3}$. As we deal with the integral effect of the turbulence on the whole
communication channel, and are not interested much in local turbulent properties, we use the well-established result
for an extended channel and the von Karman model, which predicts the following phase power
spectrum~\cite{AZB97,J06}
\begin{equation} W_\varphi (f) = \vartheta\: r_0^{-5/3}\left(f^2
	+ L_0^{-2}\right)^{-11/6} \exp(-l_0^2f^2),
	\label{phasestr}
\end{equation}
where
\begin{equation}
	\vartheta = \frac{2\sqrt{2}\G^2(11/6)}{\pi^{11/3}}\left[\frac{3}5\G(6/5)\right]^{5/6} \approx 0.0229.
\end{equation}
This shows the spectral density of optical phase $\varphi$ fluctuations with respect to the spatial frequency $f$. The
von Karman model is an empirical extrapolation of Kolmogorov's results for the whole range of spatial frequencies, as
the original theory was only applicable for the range between the inner scale $l_0$ and the outer scale $L_0$.
The parameter $r_0$ is the Fried parameter of turbulence, which tells how strong the turbulence is. One may consider it
to be the approximate diameter of the telescope, whose diffraction limit equals the turbulence-induced resolution
limit~\cite{J06}.

One reasonable approximation that we use in our analysis is that passing through a turbulent channel is equivalent to
passing through a corresponding random phase mask, whose spatial frequency spectrum is given by~(\ref{phasestr}).
Although it is not true in general (the beam may substantially re-distribute its power profile after diffraction on the
phase distortions obtained at the very beginning of the channel), it holds true for not extremely strong turbulence, where substantial
fraction of power remains in the same transverse mode. This regime is the most interesting for us,
because in the case of extremely strong turbulence, one may just assume that all output modes will be equally
populated regardless of the way they were excited at the input, which is trivial.

Phase distortion itself is a continuous function of the transverse coordinates, so we can use the Taylor series expansion to
correctly represent it.
\begin{equation}
	\varphi (x, y) = \varphi_0 + ax + by + g\frac{x^2}2 + h\frac{y^2}2 + sxy + \dots,
	\label{phase}
\end{equation}
where $a$ and $b$ are first order phase distortions, and $g$, $h$, and $s$ --- the second
order ones. They can be found as 
\begin{equation}
	\begin{array}{lll}
		a = \frac{\partial\varphi}{\partial x} & b = \frac{\partial\varphi}{\partial y}\\
		g = \frac{\partial^2\varphi}{\partial x^2} & h = \frac{\partial^2\varphi}{\partial y^2} & s =
		\frac{\partial^2\varphi}{\partial x\partial y}.
	\end{array}\label{derivatives}
\end{equation}
As phase distortion is a random function, all the mentioned distortion coefficients are random variables with
zero mean due to the apparent symmetry.

We are now ready to find the dispersion of the distortion coefficients. First, using~(\ref{derivatives}) we can find
power spectra of the distortion coefficients:
\begin{equation}\begin{array}{l}
	W_a = (2\pi)^2 f_x^2 W_\varphi\\
	W_b = (2\pi)^2 f_y^2 W_\varphi\\
	W_g = (2\pi)^4 f_x^4 W_\varphi\\
	W_h = (2\pi)^4 f_y^4 W_\varphi\\
	W_s = (2\pi)^4 f_x^2 f_y^2 W_\varphi.
\end{array}\label{distortion}\end{equation}
Where $f_x$ and $f_y$ are $x-$ and $y-$ components of the spatial frequency $f$: $f_x = f\cos(\theta)$, $f_y =
f\sin(\theta)$, where $\theta$ is the polar angle.
Second, we need to take into account that we are interested not only in the point (0,0) where we take the
derivatives~(\ref{derivatives}), but in the average phase slopes over the whole beam area. That leads to the additional
filtering function $|F(f)|^2$ similar to the one that appears in the problem of finding angle of arrival
fluctuations~\cite{BMZ92,AZB97}. In our case $|F(f)|^2$ is the spatial power spectrum of the particular mode in question.
Finally, we use the Wiener–-Khinchin theorem to find the autocorrelation of the distortion coefficients, which is a
Fourier transform of their power spectra. We may right away ignore the $x, y$-dependent part of the autocorrelation
function and only find its value for $x=y=0$, which is exactly the dispersion of the corresponding coefficients. Finding the
required zero coordinate Fourier coefficient and taking the integral over $\theta$, we obtain 
\begin{equation}
	C_a = C_b = 4\pi^3 \int_0^\infty f^3 W_\varphi (f) |F(f)|^2\,df
	\label{dispersion1}
\end{equation}
\begin{equation}
	C_g = C_h = 12\pi^5 \int_0^\infty f^5 W_\varphi (f) |F(f)|^2\,df
\end{equation}
\begin{equation}
	C_s = 4\pi^5 \int_0^\infty f^5 W_\varphi (f) |F(f)|^2\,df,
	\label{dispersions}
\end{equation}
where $C_\zeta(r) = \E[ \zeta(r_0)\zeta(r_0+r)]$ is the autocorrelation function of $\zeta$, and $C_\zeta = C_\zeta(0)$
is effectively the dispersion of $\zeta$.

Similar to~\cite{VSV12}, we assume that all the distortion coefficients are normally distributed random variables,
as they are the net effect of many independent perturbations along the optical path.
Following the same procedure one can find statistical properties of higher order distortion coefficients. In the present
paper we focus only on the first two orders because most of studied effects can be quite accurately described in this
approximation.

\section{First order approximation --- transmittance density function}
Now we calculate the probability density function of transmittance for the fundamental mode using the first-order
perturbations only.
We start with the Gaussian beam of the form
\begin{equation}
	E_{00} = \frac1w \sqrt{\frac2\pi} e^{-\frac{x^2+y^2}{w^2}},
\end{equation}
where $w$ is the beam waist. Here we assume that the channel length is not much larger than the Rayleigh range so the
beam size remains roughly the same. Calculation of the overlap integral 
\begin{equation}
	T_{00\rightarrow 00}=\frac{\left|\int  |E_{00}|^2\,
	e^{i\varphi(x,y)}\,dx\,dy\right|^2}{\left(\int|E_{00}|^2\,dx\,dy\right)^2}
	\label{overlap0}
\end{equation}
between the original beam and the linearly distorted
phase~(\ref{phase}) one yields the following power transmittance
\begin{equation}
	T_{00\rightarrow 00} = \exp\left[ -\frac{w^2}4 \left( a^2+b^2\right)\right].
	\label{t00}
\end{equation}
Denote $\xi = \frac{w^2}4 \left( a^2+b^2\right)$, which is a dimensionless perturbation.
As $a$ and $b$ are normally distributed zero mean random variables with the dispersion of $C_a = C_b$, $\xi$ has a p.d.f. of
\begin{equation}
	p(\xi) = \frac2{w^2 C_a}\exp\left(-\frac{2\xi}{w^2 C_a}\right).
	\label{xi_distrib}
\end{equation}

To find the p.d.f. of transmittance $T = f(\xi)$ we use the standard equation
\begin{equation}
	p(T) = \frac{p(\xi)}{\left|\frac{df(\xi)}{d\xi}\right|}, \mathrm{ where }\;\; \xi = f^{-1}(T).
	\label{pdf_derivation}
\end{equation}
Substituting (\ref{t00}) into (\ref{pdf_derivation}) we obtain the final p.d.f. for the channel transmittance
\begin{equation}
	p(T) = \frac{2}{w^2 C_a} T^{\frac{2}{w^2 C_a} - 1}.
	\label{power_law}
\end{equation}

One can see that in the first approximation the obtained p.d.f. is a power function of $T$, and the higher the
turbulence the smaller the power. To compare the predicted p.d.f.'s with the experiment we made series of measurements
with a single-mode optical channel passing through a turbulent chamber, where two streams of air with the specified
temperature difference are mixed together. The measured probability distributions along with the fitted theoretical
predictions are plotted in fig.~\ref{fig_transmission}.  More details on the experimental part are found in the Appendix.
The experiment and the theory match well, except at high transmittance values, where the
first-order approximation fails due to higher order phase distortions.

\begin{figure}
\includegraphics[width=\columnwidth]{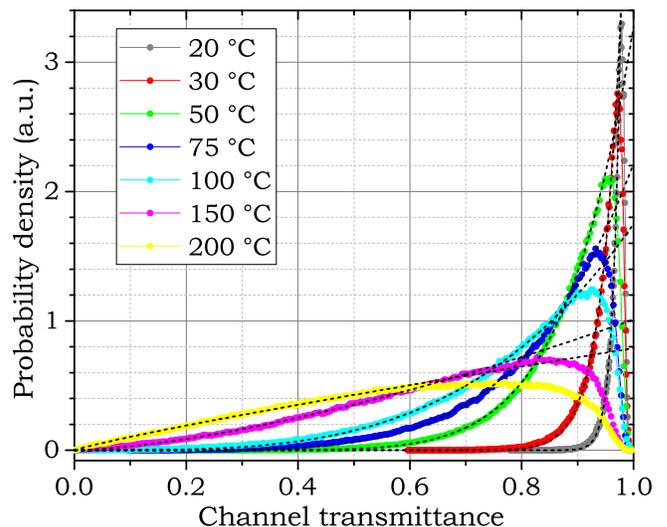}
\caption{Experimentally measured channel transmittance probability distribution and first-order theoretical
	predictions (black dashed lines). The experimental parameter governing the turbulence strength is the airflow
	temperature difference shown in the legend.
	Please note, that for the better data representation probability distributions
	are not properly normalized.}
\label{fig_transmission}
\end{figure}

\section{First order approximation --- modal cross-talk}
The next question that we address is how the power lost from the fundamental mode is distributed among higher order
modes. Here we first need to define the mode set that we use for calculations. While math is somewhat simpler for
Hermite-Gaussian modes, we wanted to find a more universal solution and succeeded by properly grouping modes together.

There is a direct correspondence between the optical mode sets (Hermite- or Laguerre-Gaussian) and a 2D isotropic
oscillator~\cite{DA92}, where the $N$-th power level is $(N+1)$-times degenerate. From the modes
perspective it means that one can group all modes according to their ``power level''. For the Hermite-Gaussian mode HG$_{mn}$
the corresponding power level is $N=m+n$. For the Laguerre-Gaussian mode LG$_{pl}$   $N=2p+|l|$. It is easy to
show that $N$-th level consists of $N+1$ distinct modes.

For each particular mode we calculate the overlap integral
\begin{equation}
	T_{00\rightarrow mn}=\frac{\left|\int E_{mn}^* E_{00}\,
	e^{i(ax+by)}\,dx\,dy\right|^2}{\int|E_{00}|^2\,dx\,dy\int|E_{mn}|^2\,dx\,dy}.
	\label{overlap_mn}
\end{equation}
After finding the integrals and grouping them by the ``power levels'' $N$, the cross-talk coefficients may be written
in terms of $\xi$ defined earlier as they lose their individual $a$ and $b$ dependence.
Here we use the explicit mode numbering for the HG set,
while it will be just different mode indices for the LG set.
For completeness, we also added the previous result for coupling back into the fundamental mode.
\begin{equation}\begin{array}{ll}
	T_0 &= T_{00\rightarrow 00} = e^{-\xi}\\
	T_1 &= T_{00\rightarrow 10,01} = \xi e^{-\xi}\\
	T_2 &= T_{00\rightarrow 20,11,02} = \frac{\xi^2}{2}e^{-\xi}\\
	T_3 &= T_{00\rightarrow 30,21,12,03} = \frac{\xi^3}{6}e^{-\xi}\\
	T_N &= T_{00\rightarrow mn: m+n=N} = \frac{\xi^N}{N!}e^{-\xi}.\\
\end{array}
\end{equation}
One can readily see that the total power is conserved as the obtained series sums up to 1.

Using~(\ref{xi_distrib}) and (\ref{pdf_derivation}) we find corresponding p.d.f.'s.
The derivative is
\begin{equation}
	\frac{df}{d\xi} = \left(1-\frac \xi N\right)\frac {\xi^{N-1}}{(N-1)!} e^{-\xi}
\end{equation}
The solution of the equation $x^N e^{-x} /N! = a$ is $x=-N\,W\left(-\frac{(a N!)^{1/N}}N\right)$, where
$W(a)$ is the Lambert $W$-function, i.e. a solution of $x e^x = a$.

The final p.d.f. of the power coupling coefficient for $N\ge 1$ is
\begin{equation}
	p(T_N) = \frac 2{w^2 C_a T} \left( \frac{\xi_1}{|N-\xi_1|} e^{-\frac{2\xi_1}{w^2 C_a}} +
	\frac{\xi_2}{|N-\xi_2|} e^{-\frac{2\xi_2}{w^2 C_a}}\right),
\end{equation}
where
\begin{equation}
	\xi_{1,2} = -N\, W\left(-\frac{(T N!)^{1/N}}N\right).
\end{equation}
The maximal power coupling for the particular mode family is $T_{N\,\mathrm{max}} = \frac{N^N}{N!} e^{-N}$.

\begin{figure}
\includegraphics[width=\columnwidth]{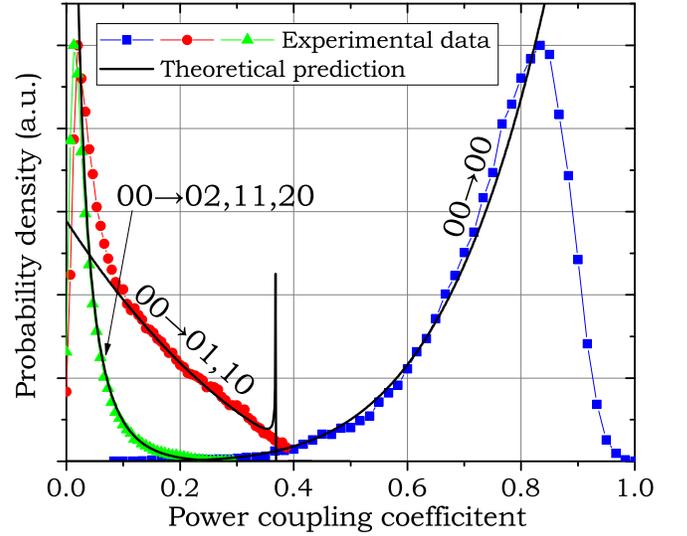}
	\caption{Experimental results and theoretical predictions of the fundamental mode coupling into the higher order modes.}
\label{fig_cross_talk_pdfs}
\end{figure}

We performed experimental measurements in the turbulent chamber with the temperature difference of $100$~$\degree$C for $N$ from 0 to 2 and show the results in fig.~\ref{fig_cross_talk_pdfs}.
The value of $C_a$ was obtained by fitting the $00\rightarrow 00$ curve by the power law~(\ref{power_law}), and the other
two curves were calculated based on this value. As in the previous case, the largest disagreement between the
theory and the experiment appears at
small values of the perturbation $\xi$, because this approximation does not take into account higher order phase
perturbations.

\section{Second and higher order approximations}
Linear phase distortions studied earlier, provide the first non-vanishing term
in the power coupling efficiency. However, this effect alone poorly describes the predicted p.d.f. at small
perturbations $\xi$,
as higher order terms start to dominate. In the following section we include quadratic terms into the
phase distortion function and calculate the corrected probability density of the fundamental mode transmittance.

\begin{figure}
\includegraphics[width=\columnwidth]{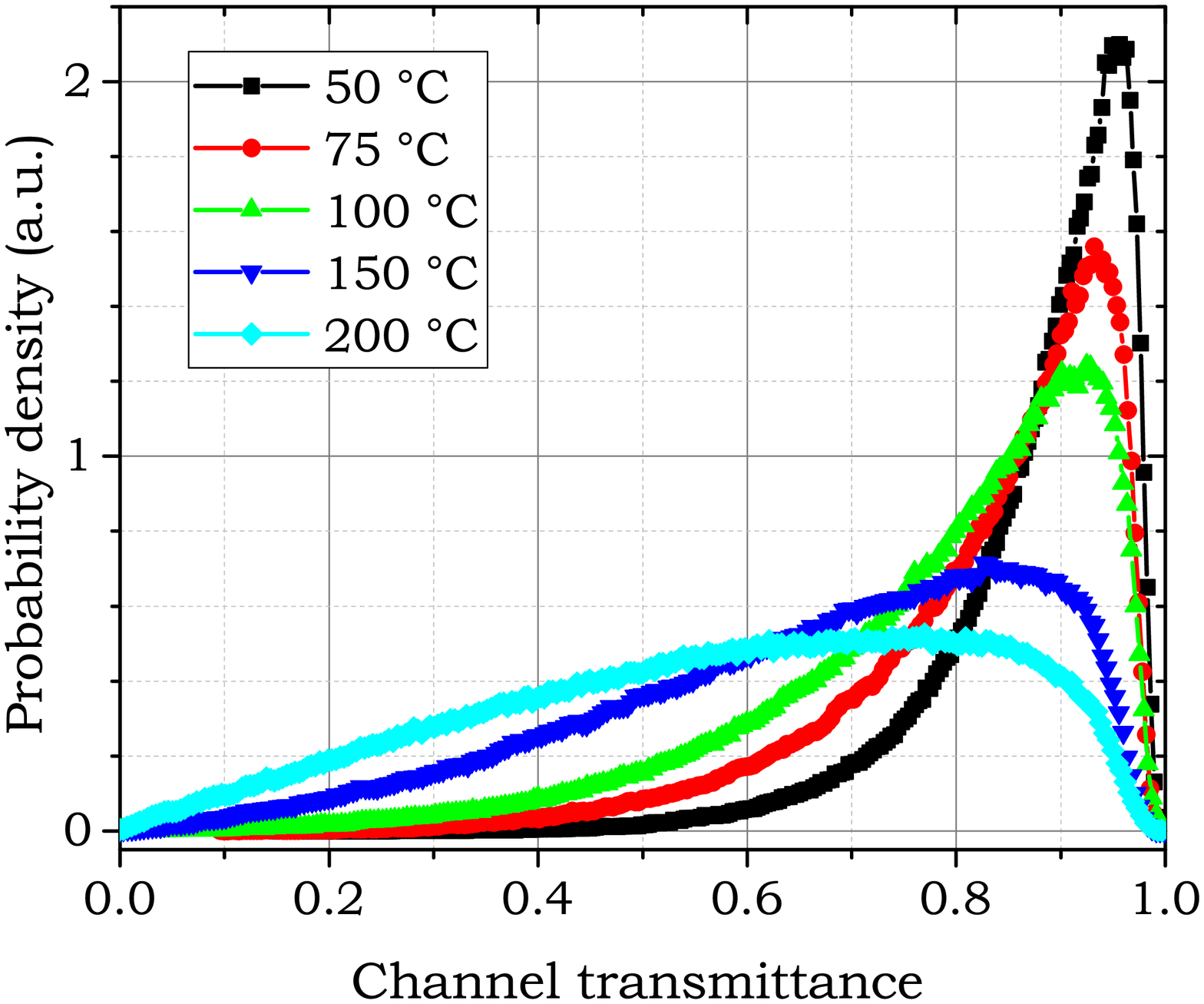}
\includegraphics[width=\columnwidth]{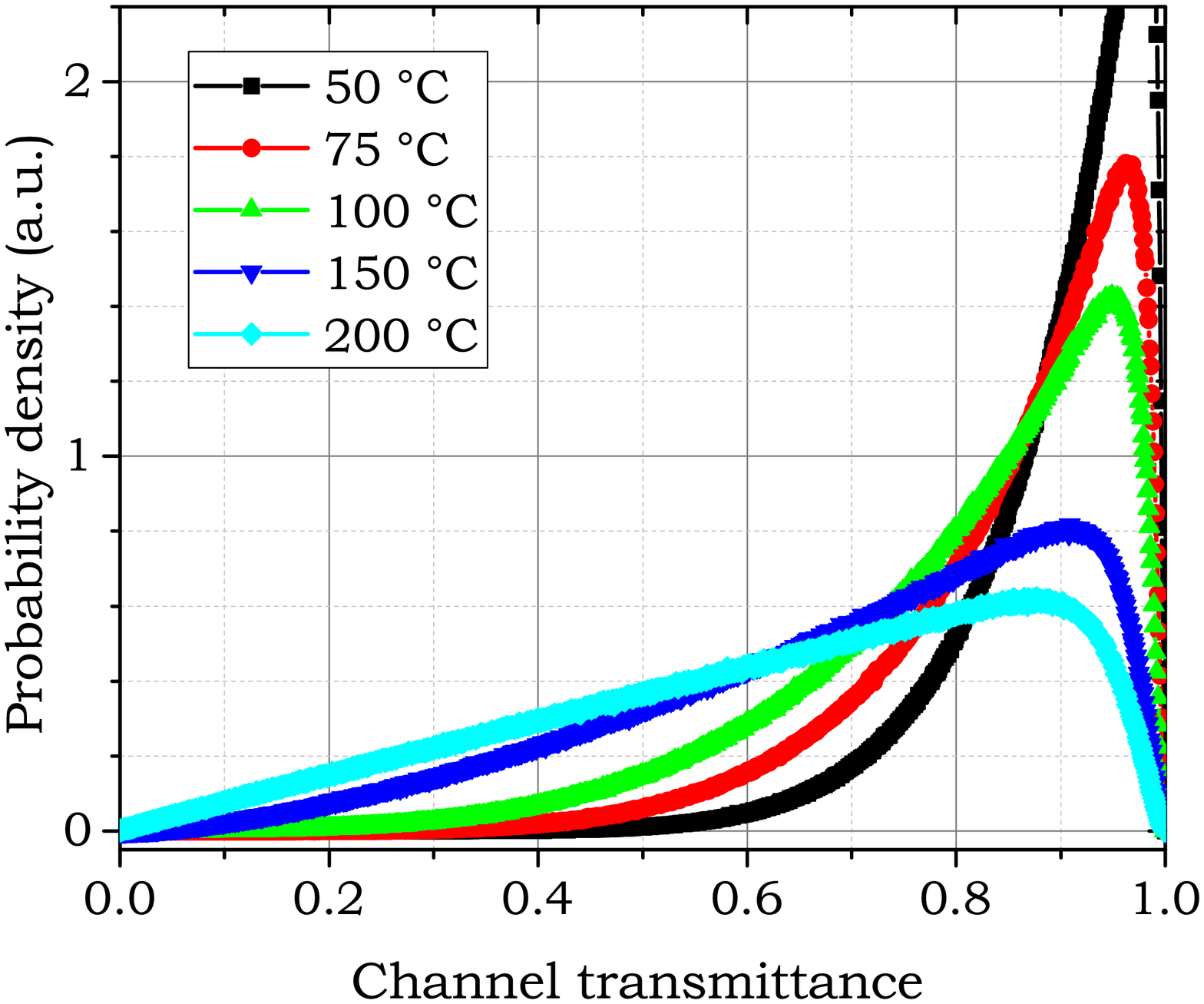}
\caption{Transmittance probability density for the fundamental mode: a comparison between the experimentally measured data
	(top) and the second-order theory predictions (bottom).}
\label{fig_quad}
\end{figure}

With the quadratic terms included the overlap integral~(\ref{overlap0}) yields
\begin{equation}
	\begin{array}{l}
		T =  \left(1+\frac{w^4}{16}(g^2+h^2+2s^2)+\frac{w^8}{256}(s^2-gh)^2\right)^{-1/2}\\
		\times \exp\left[-\frac{w^2}{16}\frac{4(a^2+b^2)+\frac{w^4}{4}(s^2a^2+s^2b^2+a^2h^2+b^2g^2-2absg -
		2absh)}{1+\frac{w^4}{16}(g^2+h^2+2s^2)+\frac{w^8}{256}(s^2-gh)^2}\right].
	\end{array}
\end{equation}
Unfortunately, analytic expressions for the transmittance probability density are unlikely to be found, so we used
numerical simulation to find the desired distributions. Again we compared the experimental data
with the results of simulations. Unlike the previous sections, where there was only one turbulence parameter
$C_a$, here we need as well to calculate $C_g$ and $C_s$. For that we relied upon the independently measured inner and outer
turbulence scales $l_0$ and $L_0$, and slightly adjusted the known Fried parameter $r_0$ to match the power of the
p.d.f.'s~(\ref{power_law}). It was necessary because the precision of independently measured $r_0$ of around 10\%
was not high enough to precisely match the expected power law fits. Based on the found turbulence parameters we
calculated $C_g$ and $C_s$ and performed the numerical simulation. The results are shown in fig.~\ref{fig_quad}.
There is a reasonably good agreement between the two, so the
presented theoretical model may be used for estimation of various derived properties of the single-mode channel.

So far we only studied the transmittance of the fundamental-mode-based free-space channel in the first and the second order
approximation as well as the cross-talk between the fundamental and higher order modes in the first approximation. This
was of the most interest because of the obtained analytic expressions that may be used for rough parameter estimation.
However, the presented framework allows to get results for any modes and precisions of phase distortion approximation.

For the particular order of phase distortion approximation one needs to find corresponding dispersions as shown
in~(\ref{distortion} -- \ref{dispersions}). Then one can construct statistically correct phase distortion functions~(\ref{phase}) and
calculate the overlap integral~(\ref{overlap_mn}) for the modes in question. Repeating this many times
one may get the desired probability distributions. Based on our measurements in the turbulent chamber, the presented
theory gives reasonable results, matching well the experimentally measured values.

\section{Discussion}
The presented framework for calculations is based solely on the well respected von Karman turbulence model,
which found many applications in predicting the results of many free-space optical communication experiments and
astronomical observations~\cite{IBD16}. So regardless on the particular experimental realization, the obtained
results are one more step towards understanding turbulent effects in single-mode optical channels and mode-multiplexed
systems.

One related practical example is implementation of an active tracking system in a single-mode free-space channel. It is
well-known that the major turbulent effect is beam wandering~\cite{VSV12}, i.e. the first order phase
distortion, while higher order effects that change the beam profile may be much weaker. At the same time, a simple
feedback loop with a fast steering mirror that controls the beam direction solves the problem of pointing error, provided the round trip time is
much shorter than the characteristic time scale of the turbulent process. As this is almost always the case, active
tracking systems substantially improve the quality of free-space optical channels, especially those delivering radiation into a
single-mode fiber~\cite{A07,A12}.

\begin{figure}
\includegraphics[width=\columnwidth]{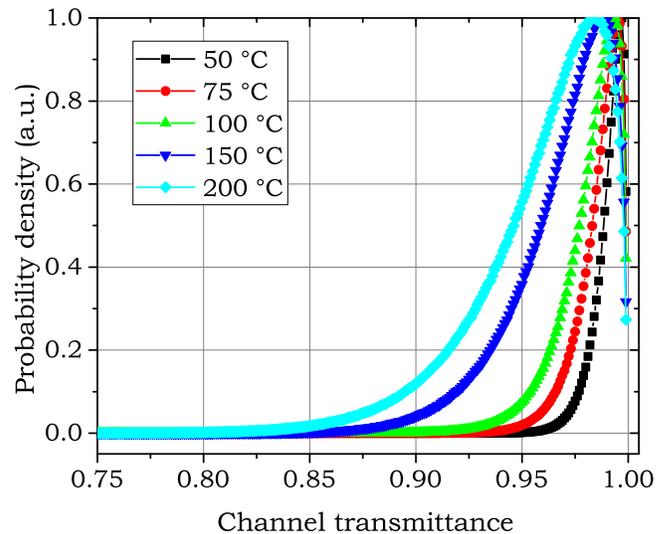}
	\caption{Simulated transmittance probability density for a free-space channel from fig.~\ref{fig_quad} with an ideal
	channel tracking implemented. The average transmittance becomes better than 95\%, which is much superior to that
	without the active tracking system. }
\label{fig_tracking}
\end{figure}

Using the developed calculation framework, one can easily estimate the channel performance provided an ideal tracking
system is implemented. To do this, the numerical simulation from the previous section is modified such that the
first-order errors $a$ and $b$ are always equal to zero. Results of such simulations are shown in
fig.~\ref{fig_tracking}, where a substantial improvement of channel performance is observed.

Another example is estimation of the Fried parameter based on the transmittance statistics for a static single-mode
channel. Measured transmittance statistics is fitted with the power function and the value $C_a$ is obtained. To
estimate the Fried parameter one uses~(\ref{dispersion1}) and a-priori knowledge of the inner and outer scales of
turbulence. In real atmosphere there are more or less known values of $l_0$ and $L_0$~\cite{J06}, while for turbulent chambers
$L_0$ is often the same as the size of the chamber, and $l_0$ is 2 -- 6 mm~\cite{KLB06}. In any case, $C_a$ weakly
depends on $l_0$ and $L_0$, and the major contribution is from the Fried parameter $r_0$. The resulting value of Fried
parameter in our experiments was always within 10\% of the independently measured one, so the described method gives
reliable results.

\section{Conclusion}
We presented a calculation framework that allows to answer most of questions regarding the performance of free-space single-mode
or mode-multiplexed channels in turbulent air.
Many first order approximations give simple analytic results that are convenient to use for quick parameter
estimation. Analytic expressions are obtained for the fundamental mode power loss and for cross-coupling between the
fundamental and higher order modes. Numerical calculations are required for more precise channel modeling that include
second and higher order phase distortions.
Many of the obtained theoretical results are supported by the experimental measurements in a turbulent chamber. Overall,
there is a good match between the experiment and calculations, which is expected provided that the von Karman turbulence
model matches the real-life environment.

\section*{Acknowledgments}
This work was partially supported by the RFBR grant No. 17-02-00966.
Results of this research work are obtained as a part of the implementation of the state support Program of NTI Centers
on the basis of educational and scientific organizations in accordance with the Rules for Granting Subsidies.

\section*{Appendix: Experimental details}
Here we briefly describe the experimental tools used for the measurements. Our turbulent chamber is based on two
5x5~cm$^2$ aluminum nozzle arrays that create jets of air in the opposite directions. The distance between the arrays is
5~cm, and one of them may be heated up to create the desired temperature difference. To calibrate the system we
measured turbulence parameters using a Shack-Hartmann wavefront sensor. We found that the inner and outer scales of
turbulence are roughly constant regardless of the temperature difference and the speed of the airflow. Their values are
$l_0 = 2.7 \pm 1 $~mm and $L_0 = 51\pm 11$~mm, respectively. The Fried parameter depends almost exclusively on the temperature difference,
getting little influence from changing the speed of airflow. This behavior seems to be common for turbulent
chambers~\cite{J06}.

In the optical part the 780-HP single-mode fiber was used for mode filtering with $F=11$~mm collimators for beam forming.
As a mode converter we used a liquid crystal based spatial light modulator (SLM) with sawtooth-like computer generated
patterns~\cite{BBS13}. The fiber-coupled optical power was converted to the electrical signal with an amplified photodiode and then
digitized at a sample rate of 1000~Hz with a universal data acquisition board.

\end{document}